\documentclass[11pt,english]{revtex4-2}
\usepackage[T1]{fontenc}
\usepackage[latin9]{inputenc}
\usepackage[letterpaper]{geometry}
\usepackage{amsmath}
\usepackage{amssymb}
\usepackage{graphicx}

\makeatletter

\providecommand{\tabularnewline}{\\}

\@ifundefined{textcolor}{}
{%
 \definecolor{BLACK}{gray}{0}
 \definecolor{WHITE}{gray}{1}
 \definecolor{RED}{rgb}{1,0,0}
 \definecolor{GREEN}{rgb}{0,1,0}
 \definecolor{BLUE}{rgb}{0,0,1}
 \definecolor{CYAN}{cmyk}{1,0,0,0}
 \definecolor{MAGENTA}{cmyk}{0,1,0,0}
 \definecolor{YELLOW}{cmyk}{0,0,1,0}
}

\usepackage{babel}

\makeatother

\usepackage{babel}
\begin{document}
\global\long\def\imag{\mathrm{i}}%
\global\long\def\ket#1{|#1\rangle}%
\global\long\def\bra#1{\langle#1|}%

\title{Counting Abelian Squares for a Problem in Quantum Computing}
\author{Ryan S. Bennink}
\affiliation{Quantum Computational Science Group, Oak Ridge National Laboratory}
\begin{abstract}
In a recent work I developed a formula for efficiently calculating
the number of abelian squares of length $t+t$ over an alphabet of
size $d$, where $d$ may be very large. Here I show how the expressiveness
of a certain class of parameterized quantum circuits can be reduced
to the problem of counting abelian squares over a large alphabet,
and use the recently developed formula to efficiently calculate this
quantity.
\end{abstract}
\maketitle

\section{Introduction}

An abelian square is a word whose first half is an anagram of its
second half, for example $\mathtt{intestines}=\mathtt{intes}\cdot\mathtt{tines}$
or $\mathtt{bonbon}=\mathtt{bon}\cdot\mathtt{bon}$. Abelian squares
have been a subject of pure math research for many decades \citep{Erdos1957,Keranen1992,Iliopoulos1997,Carpi1998,Cassaigne2011,Huova2012,Crochemore2013}
but are seemingly not encountered often in scientific applications.
Here I describe an application of abelian squares to a problem in
the field of quantum computing. This application motivated the development
of a new, more efficient recursive formula for calculating the number
of abelian squares of given length over an alphabet of given size
\citep{Bennink2022}. This work highlights the sometimes surprising
connections between pure math and applied science, and the value of
efficiently computable formulas for practitioners in applied fields.

In the first part of this article I review the basics of enumerating
abelian squares and the recently developed formula for efficiently
calculating their number. In the second part I describe the problem
of quantifying the expressiveness of parameterized quantum circuits;
show how for a particular family of circuits it reduces to the problem
of counting abelian squares over an exponentially large alphabet;
and finally, utilize the new formula to quantify the expressiveness of that
family of circuits.

\section{Counting Abelian Squares}

Let $f_{d}(t)$ denote the number of abelian squares of length $t+t$
over an alphabet of $d$ symbols. Trivially, $f_{1}(t)=1$ for all
$t$ and $f_{d}(0)=1$ for all $d$. It is also not difficult to see
that $f_{d}(1)=d$. To determine $f_{d}(t)$ for arbitrary $d$ and
$t$, we define the \emph{signature} of a word $w\in\{a_{1},\ldots,a_{d}\}^{*}$
as $(m_{1},\ldots,m_{d})$ where $m_{i}$ is the number of times the
symbol $a_{i}$ appears in $w$. Note that two words are anagrams
if and only if they have the same signature. Thus the number of abelian
squares is the number of pairs $(x,y)$ such that $x$ and $y$ have
the same signature. The number of words with a particular signature
$(m_{1},\ldots,m_{d})$ is given by the multinomial coefficient 
\begin{align}
\binom{m_{1}+\cdots+m_{d}}{m_{1},\ldots,m_{d}} & =\frac{(m_{1}+\cdots+m_{d})!}{m_{1}!\cdots m_{d}!}.
\end{align}
The number of ways to choose a pair of words, each with signature
$(m_{1},\ldots,m_{d})$, is just the square of this quantity. Therefore
the number of abelian squares of length $t+t$ is 
\begin{align}
f_{d}(t) & =\sum_{m_{1}+\cdots+m_{d}=t}\binom{t}{m_{1},\ldots,m_{d}}^{2}\label{eq: basic formula for AS count}
\end{align}
where the sum is implicitly over nonnegative integers. The first few
values of $f_{d}(t)$ are shown in Table \ref{tab: first few values}.

\begin{table}
\begin{centering}
\begin{tabular}{|c||c|c|c|c|c|c|c|c|}
\hline 
$d$\textbackslash$t$  & 0  & 1  & 2  & 3  & 4  & 5  & 6  & 7\tabularnewline
\hline 
\hline 
1  & 1  & 1  & 1  & 1  & 1  & 1  & 1  & 1\tabularnewline
\hline 
2  & 1  & 2  & 6  & 20  & 70  & 252  & 924  & 3432\tabularnewline
\hline 
3  & 1  & 3  & 15  & 93  & 639  & 4653  & 35169  & 272835\tabularnewline
\hline 
4  & 1  & 4  & 28  & 256  & 2716  & 31504  & 387136  & 4951552\tabularnewline
\hline 
5  & 1  & 5  & 45  & 545  & 7885  & 127905  & 2241225  & 41467725\tabularnewline
\hline 
6  & 1  & 6  & 66  & 996  & 18306  & 384156  & 8848236  & 218040696\tabularnewline
\hline 
\end{tabular}
\par\end{centering}
\caption{\label{tab: first few values}Number of abelian squares of length
$t+t$ over an alphabet of size $d$ \citep{Richmond2009}.}
\end{table}

Eq.~(\ref{eq: basic formula for AS count}) is not easy to evaluate
when $t$ and/or $d$ are large, as the number of signatures grows
combinatorially in $d$ and $t$. For the application to be described
in the next section, $d$ is exponentially large, prompting the need
for an efficient way of calculating $f_{d}(t)$. In \citep{Bennink2022}
I derived the recursive formula 
\begin{align}
f_{d}(t) & =d\sum_{k=0}^{t-1}\binom{t}{k}\binom{t-1}{k}f_{d-1}(k).\label{eq: new result}
\end{align}
Importantly, each level of recursion decreases both $d$ and $t$.
Thus only $\min(t,d)$ levels of recursion are needed and the cost
of evaluating $f_{d}(t)$ with this formula is only $O(t^{2}\min(d,t))$.
The fact that eq. (\ref{eq: new result}) can be evaluated efficiently
even when $d$ is exponentially large is crucial to addressing the
application described in the section.

\section{Application to a Problem in Quantum Computing}

\subsection{Parameterized Quantum Circuits and Expressiveness}

In this section I present an application of formula (\ref{eq: new result})
to a problem in the field of quantum computing. Quantum computing
is an emerging approach to computing that leverages the peculiar laws
of quantum physics to process information in new, sometimes powerful
ways. In the last few years primitive quantum computing devices have
become widely available and catapulted quantum computing into a highly
active field of research. In the current era of small, noisy devices,
the variational approach to quantum computing has become popular \citep{McClean2016,Yuan2019,Magann2021}.
In the variational approach a conventional (digital) computer adjusts
the parameters of a parameterized quantum circuit to optimize some
function of its output. This approach can be used for a variety of
useful tasks such as calculating properties of molecules and materials
\citep{Peruzzo2014,Li2017a,Kandala2017,Verdon2019,McArdle2019,Liu2021b,Kokail2019,Grimsley2019,Cirstoiu2020,Gard2020,Xu2020a,Chowdhury2020},
discrete optimization \citep{Farhi2014,Moll2018}, and machine learning
\citep{Havlicek2019,Wiebe2019,Schuld2021a,Beckey2022a,Kardashin2021},
as well as linear algebra \citep{Wang2021a} and differential equations
\citep{Lubasch2020}.

A key property of a parameterized quantum circuit is its expressiveness---the
range of outputs that can be obtained by varying the parameters. A
circuit that is not expressive enough for the problem at hand will
produce inferior solutions. On the other hand, a circuit that is overly
expressive may be difficult to optimize \citep{Holmes2022}. For our
purposes, the output of a quantum circuit will be the state of an
$n$-qubit register. (A qubit is a quantum bit.) Such a state can
be represented by a unit-length complex vector $\psi\in\mathbb{C}^{2^{n}}$,
with the caveat that the overall complex phase of the state is irrelevant.

One way of quantifying the expressiveness of a parameterized circuit
is by its fidelity distribution \citep{Sim2019,Rasmussen2020}. Fidelity
$F(\psi,\psi^{\prime})=\left|\left<\psi,\psi^{\prime}\right>\right|^{2}$,
where $\langle\cdot,\cdot\rangle$ denotes inner product, is a measure
of the similarity of two quantum states $\psi$ and $\psi^{\prime}$.
It ranges from 0 (for completely dissimilar states) to 1 (for identical
states). Let $\psi(\theta)$ denote the quantum state produced by
a quantum circuit as a function of the parameter vector $\theta$.
Suppose parameter values are drawn at random. If the circuit is highly
expressive, i.e. capable of producing a wide range of states, most
of the resulting states will be dissimilar to each other and will
have small mutual fidelity. Conversely, if the circuit is inexpressive,
i.e. capable of producing only a narrow range of states, most of the
produced states will be similar to each other and have large mutual
fidelity. Thus the expected value of $F(\psi(\theta),\psi(\theta^{\prime}))$,
where $\theta,\theta^{\prime}$ are independent random parameter values,
quantifies the circuit's expressiveness: the lower the expected value,
the more expressive the circuit. As it turns out, this metric is not
very sensitive. A more discerning metric is 
\begin{equation}
\mathbb{E}\left[F(\psi(\theta),\psi(\theta^{\prime}))^{t}\right]
\end{equation}
where $t>1$; typically $t$ is a small positive integer. As $t$
increases, $\mathbb{E}\left[F^{t}\right]$ becomes less sensitive
to the states that are far apart. Thus small values of $t$ measure
the expressiveness at a coarse scale in the quantum state space, while
large values of $t$ measure the expressiveness at a fine scale.

\subsection{Commutative Quantum Circuits and Abelian Squares}

Commutative quantum circuits (also known as Instantaneous Quantum
Polynomial circuits \citep{Shepherd2009}) are a class of relatively
simple parameterized quantum circuits whose output distributions are
hard to simulate using digital computers \citep{Bremner2011}. These
properties make them an interesting case study in the quest to understand
when and why quantum computing is more powerful than classical computing.
These properties also suggests that commutative quantum circuits may
be a useful ansatz for variational quantum algorithms, for example
in the field of machine learning \citep{Coyle2020a}.

\begin{figure}
\begin{centering}
{\Large{}\includegraphics[width=3in]{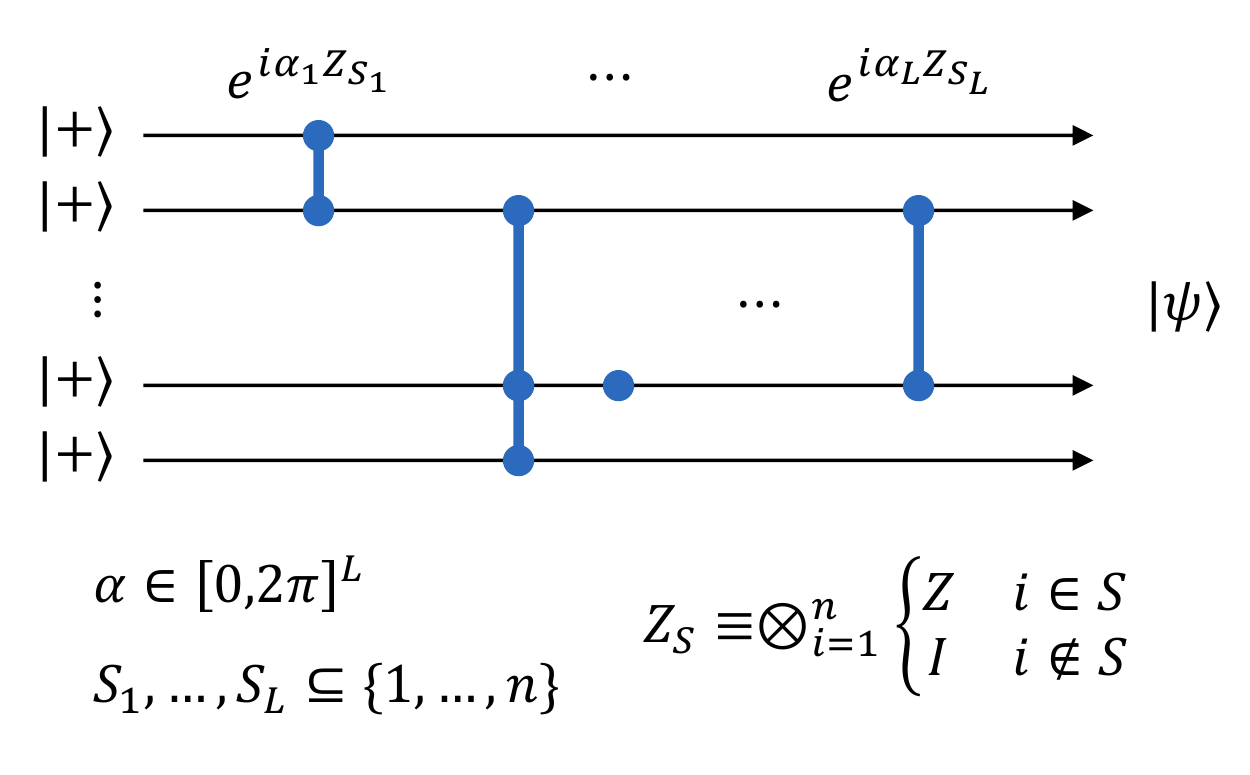}}{\Large\par}
\par\end{centering}
\begin{centering}
 
\par\end{centering}
\caption{\label{fig: circuit}The structure of a commutative quantum circuit
(CQC) in the Hadamard basis. Each circuit operation is a multiqubit
$Z$-type rotation on a distinct subset of qubits. A maximal CQC on
$n$ qubits consists of all $2^{n}-1$ $Z$-type rotations that act
non-trivially on at least one qubit.}
\end{figure}

An $n$ qubit commutative quantum circuit (CQC) of length $L$ can
be defined as a sequence of $L$ multiqubit $X$ rotations acting
on the state $\ket 0^{\otimes n}$. (Since these operations all commute,
their order does not matter.) For our purposes it will be convenient
to treat the circuit and its output in the Hadamard basis; in this
basis the circuit consists of $L$ multiqubit $Z$ rotations acting
on the state $\ket +^{\otimes n}$ where $\ket +\equiv(\ket 0+\ket 1)/\sqrt{2}$
(Fig. \ref{fig: circuit}). The output state is 
\begin{align}
\ket{\psi} & =\left(\prod_{j=1}^{L}\exp(\imag\alpha_{j}Z_{S_{j}})\right)\sum_{x\in\{0,1\}^{n}}\frac{1}{\sqrt{2^{n}}}\ket x
\end{align}
Where $S_{1},\ldots,S_{L}$ are distinct subsets of $\{1,\ldots,n\}$
and $Z_{S}\equiv\bigotimes_{i=1}^{n}\begin{cases}
Z & i\in S\\
I & i\not\in S
\end{cases}$, with $Z\equiv\ket 0\bra 0-\ket 1\bra 1$.

Consider a ``maximal'' circuit consisting of all $2^{n}$ $Z$-type
rotations. Then $\alpha$ may be regarded as a vector over all length-$n$
bitstrings, where each bitstring specifies a particular subset of
$\{1,\ldots,n\}$. A simple derivation shows that 
\begin{align}
\ket{\psi} & =\frac{1}{\sqrt{2^{n}}}\sum_{x\in\{0,1\}^{n}}e^{\imag\theta_{x}}\ket x
\end{align}
where $\theta\in\mathbb{R}^{2^{n}}$ is the Walsh-Hadamard transform
of $\alpha$. It follows that the ability to prescribe all $2^{n}$
components of $\alpha$ implies the ability to prescribe all $d$
components of $\theta$. Since the circuit operation corresponding
to $\alpha_{0}$ imparts an inconsequential global phase to the quantum
state, that circuit operation may be omitted and the global phase
may be chosen so that $\theta_{0}=0$. The output state may then be
represented by a length-$2^{n}$ complex vector 
\begin{align}
\psi(\theta) & =\left(\frac{1}{\sqrt{d}},\frac{e^{\imag\theta_{1}}}{\sqrt{d}},\ldots,\frac{e^{\imag\theta_{d-1}}}{\sqrt{d}}\right)
\end{align}
where $\theta_{1},\ldots,\theta_{d-1}$ can be independently varied.
(Here $d=2^{n}$ and I have switched indices from bitstrings in $\{0,1\}^{n}$
to corresponding integers in $\{0,\ldots,2^{n}-1\}$.)

While maximal commutative quantum circuits are not practically realizable
for large $n$ (the number of operations is $2^{n}-1$), they provide
an upper bound on the expressiveness that can be achieved by any commutative
quantum circuit with a given number of qubits. As I will now show,
the expressiveness of a maximal commutative circuit, as measured by
$\mathbb{E}\left[F^{t}\right]$, is proportional to $f_{2^{n}}(t)$.
The fidelity $F$ is the square of the inner product 
\begin{align}
\psi(\theta)^{\dagger}\psi(\theta^{\prime}) & =\frac{1}{d}\sum_{x=0}^{d-1}e^{\imag(\theta_{x}^{\prime}-\theta_{x})}.
\end{align}
In terms of $\phi_{x}\equiv\theta_{x}^{\prime}-\theta_{x}$ we have
\begin{align}
F(\psi(\theta),\psi(\theta^{\prime})) & =\left|\psi(\theta)^{\dagger}\psi(\theta^{\prime})\right|^{2}=\frac{1}{d^{2}}\sum_{x,y=0}^{d-1}e^{\imag(\phi_{x}-\phi_{y})},
\end{align}
\begin{align}
F(\psi(\theta),\psi(\theta^{\prime}))^{t} & =\frac{1}{d^{2t}}\sum_{x_{1},y_{1}=0}^{d-1}\cdots\sum_{x_{t},y_{t}=0}^{d-1}e^{\imag(\phi_{x_{1}}+\cdots+\phi_{x_{t}})-\imag(\phi_{y_{1}}+\cdots\phi_{y_{t}})},
\end{align}
and 
\begin{align}
\mathbb{E}\left[F(\psi(\theta),\psi(\theta^{\prime}))^{t}\right] & =\frac{1}{d^{2t}}\sum_{x_{1},y_{1}=0}^{d-1}\cdots\sum_{x_{t},y_{t}=0}^{d-1}\mathbb{E}\left[e^{\imag(\phi_{x_{1}}+\cdots+\phi_{x_{t}})-\imag(\phi_{y_{1}}+\cdots\phi_{y_{t}})}\right].
\end{align}
Let us suppose the rotation angles $\alpha_{i}$ are drawn uniformly
and independently from $[0,2\pi]$. Then each $\theta_{x}$ and $\theta_{x}^{\prime}$
are independent and uniform over $[0,2\pi]$, and $\phi_{x}$ is also
uniform over $[0,2\pi]$. For each $i\in\{1,\ldots,d-1\}$, let $m_{i}(x)$
be the number of occurrences of $i$ in $(x_{1},\ldots,x_{t})$ and
let $m_{i}(y)$ denote the number of occurrences of $i$ in $(y_{1},\ldots,y_{t})$.
Then the summand may be written as 
\begin{align}
\mathbb{E}\left[e^{\imag(\phi_{x_{1}}+\cdots+\phi_{x_{t}})-\imag(\phi_{y_{1}}+\cdots\phi_{y_{t}})}\right] & =\mathbb{E}\left[\prod_{i=1}^{d-1}e^{\imag(m_{i}(x)-m_{i}(y))\phi_{i}}\right]\\
 & =\prod_{i=1}^{d-1}\mathbb{E}\left[e^{\imag(m_{i}(x)-m_{i}(y))\phi_{i}}\right]
\end{align}
since the $\phi_{i}$'s are independent. Now, 
\begin{align}
\mathbb{E}\left[e^{\imag(m_{i}(x)-m_{i}(y))\phi_{i}}\right] & =\begin{cases}
1 & m_{i}(x)=m_{i}(y)\\
0 & m_{i}(x)\ne m_{i}(y)
\end{cases}.
\end{align}
Thus the only pairs $(x,y)$ that contribute to $\mathbb{E}\left[F^{t}\right]$
are those for which $m_{i}(x)=m_{i}(y)$ for all $i=1,\ldots,d-1$.
For such pairs it also holds that $m_{0}(x)=m_{0}(y)$. That is, a
term contributes if and only if $x=(x_{1},\ldots,x_{t})$ is an anagram
of $y=(y_{1},\ldots,y_{t})$, i.e. $xy$ is an abeliean square. It
follows that 
\begin{align}
\mathbb{E}\left[F^{t}\right] & =\frac{f_{d}(t)}{4^{nt}}.
\end{align}
Whereas $t$ is typically small, $d=2^{n}$ can be very large, which
necessitates use of eq. (\ref{eq: new result}).

\begin{figure}
\begin{centering}
\includegraphics[width=2.5in]{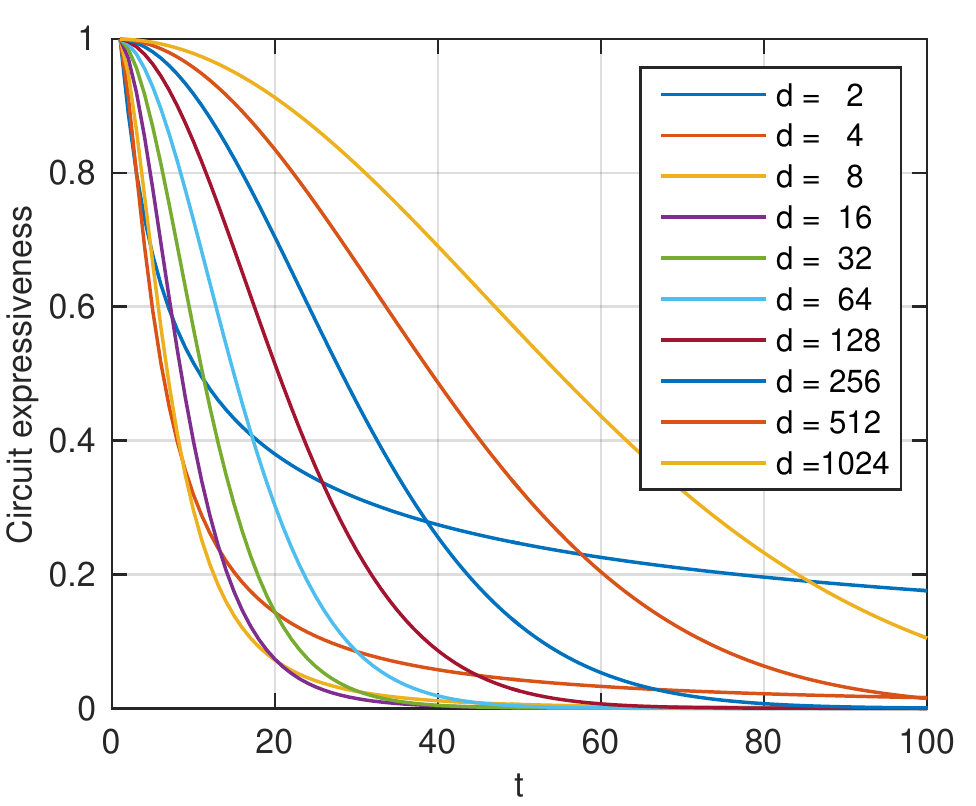} 
\par\end{centering}
\caption{\label{fig: expressivness plots}Normalized expressiveness $\mathbb{E}\left[F^{t}\right]_{\text{min}}/\mathbb{E}\left[F^{t}\right]$
of maximal commutative quantum circuits, as a function of the number
of qubits $n$ and the resolving power $t$. }
\end{figure}

It is convenient to compare $\mathbb{E}\left[F^{t}\right]$ for a
given circuit to its minimal value 
\begin{align}
\mathbb{E}\left[F^{t}\right]_{\text{min}} & =\binom{t+d-1}{t}^{-1}
\end{align}
which is achieved by a circuit that covers the entire state space
uniformly. Fig. \ref{fig: expressivness plots} plots the normalized
expressiveness $\mathbb{E}\left[F^{t}\right]_{\text{min}}/\mathbb{E}\left[F^{t}\right]$.
For all $d$, the normalized expressive is near 1 at small $t$ and
decays to 0 at large $t$. This indicates that the circuits are highly
expressive at coarse scales (small $t$), but have very low expressiveness
at fine scales (large $t$). That is, the set of states that can achieved
by maxmimal commutative quantum circuits span the breadth of the state
space, but constitute only a sparse or low-dimensional subset of the
state space.

\section{Acknowledgements}

This work was performed at Oak Ridge National Laboratory, operated
by UT-Battelle, LLC under contract DE-AC05-00OR22725 for the US Department
of Energy (DOE). Support for the work came from the DOE Advanced Scientific
Computing Research (ASCR) Accelerated Research in Quantum Computing
Program under field work proposal ERKJ354.


\end{document}